\documentstyle[psfig]{sup}
\vspace{-8pt}

\title[Superlattices and Microstructures, Vol.\ --, No.\ -,
1998] {Flux Confinement in Mesoscopic
Superconductors}
\author[Superlattices and Microstructures,
Vol.\ --, No.\ -, 1998]{Y.~Bruynseraede
and V.~V.~Moshchalkov%
\cr\vspace{10pt}%
{\normalsize\it Laboratorium voor Vaste-Stoffysica
en Magnetisme, K.U.Leuven, B-3001 Leuven,
Belgium}\cr}

\pagerange{\pageref{firstpage}--\pageref{lastpage}}

\def\LaTeX{L\kern-.25em\raise.425ex\hbox{a}\kern-.075em\TeX}

\def\fakebold#1{\relax\ifvmode\leavevmode\fi%
\ifmmode%
\setbox0=\hbox{$#1$}%
\else%
\setbox0=\hbox{#1}%
\fi%
\kern-.02em\copy0 \kern-\wd0%
\kern .04em\copy0 \kern-\wd0%
\kern-.0125em\raise.02em\box0%
}%
\begin{document}
\label{firstpage}
\maketitle
\sloppy
\begin{center}
\received{(Received 9 November 1998)}
\end{center}
%**********************************************************************%
%
\begin{abstract}
We report on flux confinement effects in
superconducting submicron line, loop and dot
structures. The main idea of our study was to vary
the boundary conditions for confinement of the
superconducting condensate by taking samples of
different topology and, through that, modifying the
lowest Landau level $E_{LLL}(H)$. Since the critical
temperature versus applied magnetic field $T_{c}(H)$
is, in fact, $E_{LLL}(H)$ measured in temperature
units, it is varied as well when the sample topology
is changed. We demonstrate that in all studied
submicron structures the shape of the $T_{c}(H)$
phase boundary is determined by the confinement
topology in a unique way.
\end{abstract}

\section{Introduction}

Recent impressive progress in nanostructuring
(e-beam lithography, single atom manipulation with a
scanning tunneling microscope tip, etc.) has made it
possible to control the quantization effects in
nanostructures by varying their size and topology.
The most commonly used so far for this purpose are
semiconducting structures, where an elegant approach
developed by Landauer~\cite{Landauer} relates
directly mesoscopic transport to the quantum
transition probability, which is determined by the
specific configuration of quantum levels and
available tunneling barriers. Changing the
confinement potential via nanostructuring allows to
tune both the quantum levels and the tunneling
probabilities. Extended by
B\"{u}ttiker~\cite{Buttiker} to the multi-terminal
measurement geometry, the Landauer-B\"{u}ttiker
formalism has been widely and very successfully used
in the interpretation of numerous mesoscopic
experiments~\cite{Beenakker}.

In comparison to the semiconducting and normal
systems, {\it superconducting structures} have been
studied much less so far. In this report, it is
worth first to ask a few simple questions like: why
do we want to make such structures, what interesting
new physics do we expect, and why do we want to
focus on superconducting (and not, for example,
normal metallic) nanostructured materials?

First of all, by making low dimensional systems, one
creates an artificial potential in which charge
carriers or flux lines are confined. The confinement
length scale $L_{A}$ of an elementary "plaquette" A,
gives roughly the expected energy scale $E=\hbar^2
\pi^2 n^2/(2 m L_A^2)$. The concentration of charge
carriers or flux lines can be controlled by varying
the gate voltage in two-dimensional (2D) electron
gas systems~\cite{Ensslin} or the applied magnetic
field in superconductors~\cite{Pannetier84}. In this
situation, different commensurability effects
between the fixed number of elements A in an array
and a tunable number of charge or flux carriers are
observed.

Secondly, modifying the sample topology in those
systems creates a unique possibility to impose the
desired boundary conditions, and thus almost
"impose" the properties of the sample. A Fermi
liquid or a superconducting condensate confined
within such a system will be subjected to severe
constraints and, as a result, the properties of
these systems will be strongly affected by the
boundary conditions.

While a normal metallic system should be considered
quantum-mechanically by solving the Schr\"{o}dinger
equation:
\begin{equation}
\frac{1}{2 m}
\left( - \imath \hbar \vec{\nabla} - e \vec{A} \right)^{2} \Psi +U \: \Psi
= E \: \Psi \: ,
\label{e}
\end{equation}
a superconducting system is described by the two
coupled Ginzburg-Landau (GL) equations:
\begin{equation}
\frac{1}{2m^{\star}}(-i \hbar\vec{\nabla}-e^{\star}
\vec{A})^{2}\Psi_s+\beta |\Psi_s|^{2} \Psi_s = -\alpha \Psi_s
\label{GLFree1}
\end{equation}
\begin{equation}
\vec{j_{S}}=\vec{\nabla} \times \vec{h} = \frac{e^{\star}}{2 m^{\star}}
\left[ \Psi_s^{\star} (- \imath \hbar \vec{\nabla} - e^{\star} \vec{A})
\Psi_s +
\Psi_s ( \imath \hbar \vec{\nabla} - e^{\star} \vec{A}) \Psi_s^{\star}
\right] \: ,
\label{GL2A}
\end{equation}
with $\vec{A}$ the vector potential which
corresponds to the microscopic field
$\vec{h}=rot\vec{A}/\mu_{0}$, $U$ the potential
energy, $E$ the total energy, $\alpha$ a temperature
dependent parameter changing sign from $\alpha > 0$
to $\alpha < 0$ as $T$ is decreased through $T_{c}$,
${\beta}$ a positive temperature independent
constant, $m^{\star}$ the effective mass which can
be chosen arbitrarily and is generally taken as
twice the free electron mass $m$.

Note that the first GL~equation
(Eq.~(\ref{GLFree1})), with the nonlinear term
$\beta |\Psi_s|^{2} \Psi_s$ neglected, is the
analogue of the Schr\"{o}dinger equation
(Eq.~(\ref{e})) with $U$ = 0, when making the
following substitutions: $\Psi_{s}(\Psi$),
$e^{\star}(e)$, $-\alpha(E)$ and $m^{\star}(m)$. The
superconducting order parameter $\Psi_{s}$
corresponds to the wave function $\Psi$; the
effective charge $e^{\star}$ in the GL equations is
$2 e$, i.e. the charge of a Cooper pair; the
temperature dependent GL parameter $\alpha$
\begin{equation}
- \alpha = \frac{\hbar^{2}}{2 m^{\star} \: \xi^{2}(T)}
\label{GLAlpha}
\end{equation}
plays the role of $E$ in the Schr\"{o}dinger
equation. Here $\xi(T)$ is the temperature dependent
coherence length:
\begin{equation}
\xi(T)=\xi(0) \, \left( 1-\frac{T}{T_{c0}} \right)^{-1/2}.
\label{XiT}
\end{equation}

\begin{figure}[h]
\centerline{\psfig{figure=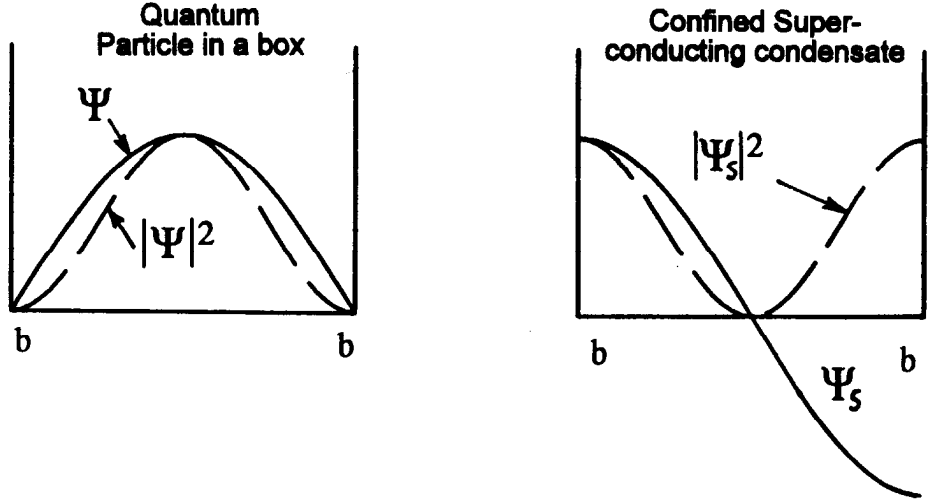}}
% \vspace{6.4cm}
\caption{Boundary conditions for interfaces between normal
metal-vacuum and
\mbox{superconductor}-vacuum.}
\label{FD}
\end{figure}

The boundary conditions for interfaces between
normal metal-vacuum and superconductor-vacuum are,
however, different (Fig.~\ref{FD}):
\begin{equation}
\left. \Psi\Psi^{\star} \right|_{b}=0
\label{EBound}
\end{equation}
\begin{equation}
\left. (- \imath \hbar \vec{\nabla} - e^{\star} \vec{A}) \Psi_{s}
\right|_{\perp , b}=0
\label{GLBound}
\end{equation}
i.e. for normal metallic systems the density is
zero, while for superconducting systems, the
gradient of $\Psi_s$ (for the case $\vec{A}=0$) has
no component perpendicular to the boundary. As a
consequence, the supercurrent cannot flow through
the boundary. The nucleation of the superconducting
condensate is favored at the superconductor/ vacuum
interfaces, thus leading to the appearance of
superconductivity in a surface sheet with a
thickness $\xi(T)$ at the third critical field
$H_{c3}(T)$.

For bulk superconductors the surface-to-volume ratio
is negligible and therefore superconductivity in the
bulk is not affected by a thin superconducting
surface layer. For submicron superconductors with
antidot arrays, however, the boundary conditions
(Eq.~(\ref{GLBound})) and the surface
superconductivity introduced through them, become
very important if $L_{A}\leq\xi(T)$. The advantage
of superconducting materials in this case is that it
is not even necessary to go down to the nanometer
scale (like for normal metals), since for $L_{A}$ of
the order of 0.1-1.0~$\mu m$ the temperature range
where $L_{A}
\leq \xi(T)$, spreads over $0.01-0.1~K$ below
$T_{c}$ due to the divergence of $\xi(T)$ at
$T\rightarrow T_{c0}$ (Eq.~(\ref{XiT})).

In principle, the mesoscopic regime
$L_{A}\leq\xi(T)$ can be reached even in bulk
superconducting samples with $L_{A}\sim$ $1~cm$,
since $\xi(T)$ diverges. However, the temperature
window where $L_{A}\leq\xi(T)$ is so narrow, not
more than $\sim$ $1~nK$ below $T_{c0}$, that one
needs ideal sample homogeneity and a perfect
temperature stability.

In the mesoscopic regime, $L_{A}\leq\xi(T)$, which
is easily realized in (perforated) nanostructured
materials, the surface superconductivity can cover
the whole available space occupied by the material,
thus spreading superconductivity all over the
sample. It is then evident that in this case surface
effects play the role of bulk effects.

Using the similarity between the linearized
GL~equation (Eq.~(\ref{GLFree1})) and the
Schr\"{o}dinger equation (Eq.~(\ref{e})), we can
formalize our approach as follows: since the
parameter
-$\alpha$ (Eqs.~(\ref{GLFree1}) and~(\ref{GLAlpha})) plays the role of
energy $E$ (Eq.~(\ref{e})), then {\it the highest
possible temperature $T_{c}(H)$ for the nucleation
of the super\-con\-duc\-ting state in the presence
of a magnetic field $H$ always corresponds to the
lowest Landau level $E_{LLL}(H)$} found by solving
the Schr\"{o}dinger equation (Eq.~(\ref{e})) with
"superconducting" boundary
conditions~(Eq.~(\ref{GLBound})).

\begin{figure}[h]
\centerline{\psfig{figure=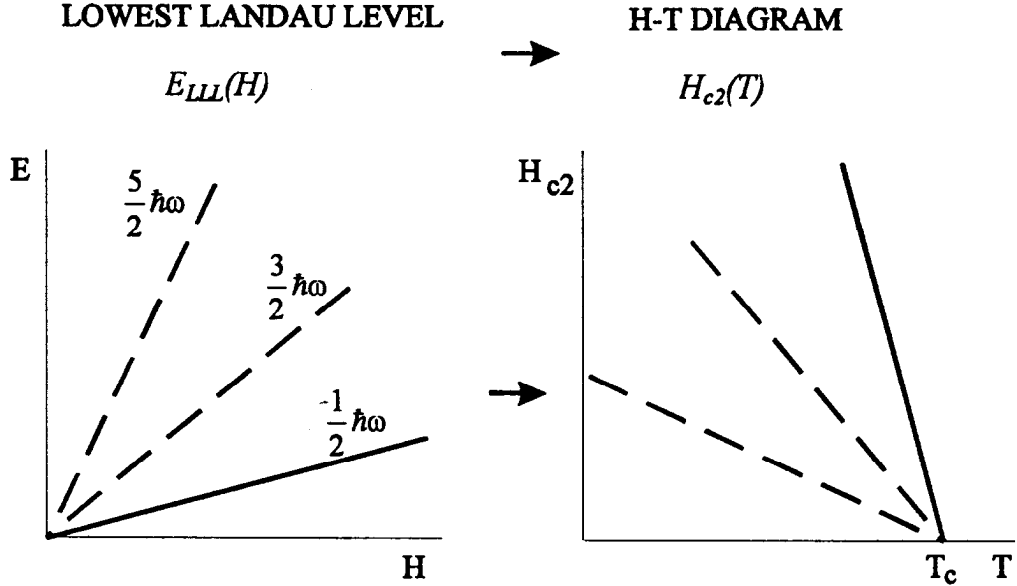}}
% \vspace{8.6cm}
\caption{Landau level scheme for a particle in a magnetic field. From the
lowest Landau level $E_{LLL}(H)$ the second critical field $H_{c2}(T)$
is derived (solid line).}
\label{FE}
\end{figure}

Figure~\ref{FE} illustrates the application of this
rule to the calculation of the upper critical field
$H_{c2}(T)$: indeed, if we take the well-known
classical Landau solution for the lowest level in
bulk samples $E_{LLL}(H)=\hbar\omega/2$, where
$\omega = e^{\star} \mu_0 H / m^{\star}$ is the
cyclotron frequency. Then, from -$\alpha =
E_{LLL}(H)$ we have
\begin{equation}
\frac{\hbar^{2}}{2 m^{\star} \: \xi^{2}(T)}=
\left. \frac{\hbar\omega}{2}\right|_{H=H_{c2}}
\label{ha}
\end{equation}
and with the help of Eq.~(\ref{GLAlpha}), we obtain
\begin{equation}
\mu_{0} H_{c2}(T)=\frac{\Phi_{0}}{2 \pi \xi^{2}(T)}
\label{hc2}
\end{equation}
with $\Phi_0 = h/e^{\star} = h/2 e$ the
superconducting flux quantum.

In nanostructured superconductors, where the
boundary conditions (Eq.~(\ref{GLBound})) strongly
influence the Landau level scheme, $E_{LLL}(H)$ has
to be calculated for each different confinement
geometry. By measuring the shift of the critical
temperature $T_c(H)$ in a magnetic field, we can
compare the experimental $T_{c}(H)$ with the
calculated level $E_{LLL}(H)$ and thus check the
effect of the confinement topology on the
superconducting phase boundary for a series of
nanostructured superconducting samples. The
transition between normal and superconducting states
is usually very sharp and therefore the lowest
Landau level can be easily traced as a function of
applied magnetic field. Except when stated
explicitly, we have taken the midpoint of the
resistive transition from the superconducting to the
normal state, as the criterion to determine
$T_c(H)$.

\section{Flux confinement in individual structures:
line, loop and dot}

In this section we present the experimental $T_c(H)$
phase boundary measured in superconducting aluminum
mesoscopic structures with different topologies with
the same width of the lines ($w=0.15
\: \mu m$) and film thickness ($t= 25 \: nm$).
The magnetic field $H$ is always applied
perpendicular to the structures.

\subsection{Line structure}
\begin{figure}[h]
\centerline{\psfig{figure=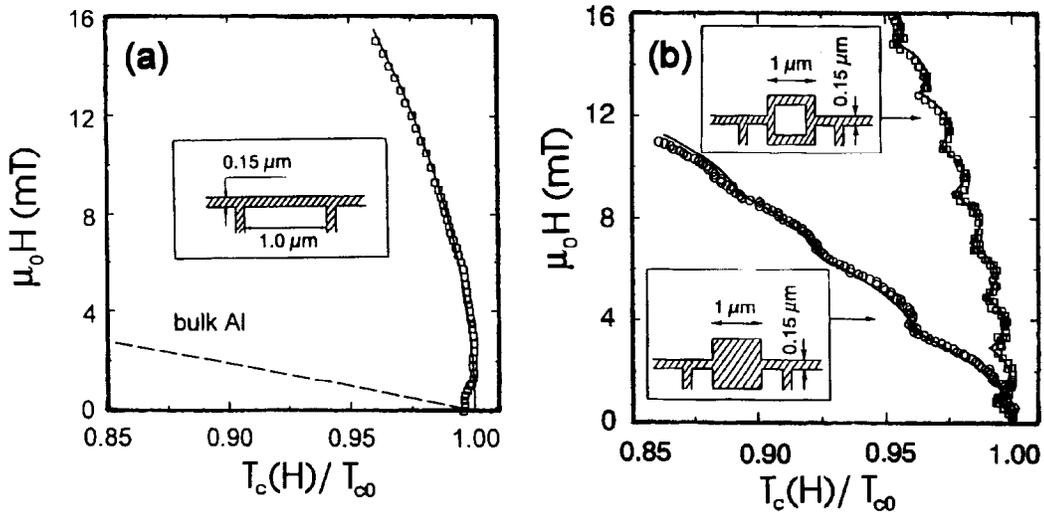}}
% \vspace{7.2cm}
\caption{The measured superconducting/normal phase boundary
as a function of the reduced temperature
$T_c(H)/T_{c0}$ for a)~the line structure, and
b)~the loop and dot structure. The solid line in (a)
is calculated using Eq.~(\ref{TCBLine}) with $\xi
(0)=110 \: nm$ as a fitting parameter. The dashed
line represents $T_c(H)$ for bulk Al.}
\label{MesLine}
\end{figure}

In Fig.~\ref{MesLine}a the phase boundary $T_c(H)$
of a mesoscopic line is shown. The solid line gives
the $T_c(H)$ calculated from the well-known
formula~\cite{Tin63}:
\begin{equation}
T_{c}(H)=T_{c0} \left[ 1 - \frac{\pi^{2}}{3}
\left( \frac{w \: \xi(0) \mu_0 H}{\Phi_0}\right)^{2} \right]
\label{TCBLine}
\end{equation}
which, in fact, describes the parabolic shape of
$T_c(H)$ for a thin film of thickness $w$ in a
parallel magnetic field. Since the cross-section,
exposed to the applied magnetic field, is the same
for a film of thickness $w$ in a parallel magnetic
field and for a mesoscopic line of width $w$ in a
perpendicular field, the same formula can be used
for both configurations~\cite{VVM95}. Indeed, the
solid line in Fig~\ref{MesLine}a is a parabolic fit
of the experimental data with Eq.~(\ref{TCBLine})
where $\xi (0)=110 \: nm$ was obtained as a fitting
parameter. The coherence length obtained using this
method, coincides reasonably well with the dirty
limit value $\xi(0)= 0.85 (\xi_0 \ell)^{1/2}= 132 \:
nm$ calculated from the known BCS coherence length
$\xi_0=1600 \: nm$ for bulk Al~\cite{dGABook} and
the mean free path $\ell = 15 \: nm$, estimated from
the normal state resistivity $\rho$ at $4.2 \:
K$~\cite{Rom82}.

We can use also another simple argument to explain
the parabolic relation $T_c(H) \propto H^2$, since
the expansion of the energy $E(H)$ in powers of $H$,
as given by the perturbation theory,
is~\cite{Wel38}:
\begin{equation}
E(H)=E_0+A_1 L H + A_2 S H^2 + \cdots
\label{Perturb}
\end{equation}
where $A_1$ and $A_2$ are constant coefficients, the
first term $E_0$ represents the energy levels in
zero field, the second term is the linear field
splitting with the orbital quantum number $L$ and
the third term is the diamagnetic shift with $S$,
being the area exposed to the applied field.

For the topology of the line with a width $w$ much
smaller than the Larmor radius $r_H \gg w$, any
orbital motion is impossible due to the constraints
imposed by the boundaries onto the electrons inside
the line. Therefore, in this particular case $L=0$
and $E(H)=E_0 + A_2 S H^2$, which immediately leads
to the parabolic relation $T_c \propto H^2$. This
diamagnetic shift of $T_c(H)$ can be understood in
terms of a partial screening of the magnetic field
$H$ due to the non-zero width of the
line~\cite{Tinkham75}.

\subsection{Loop structure}
The $T_c(H)$ of the mesoscopic loop, shown in
Fig.~\ref{MesLine}b, demonstrates very distinct
Little-Parks (LP) oscillations~\cite{Litt62}
superimposed on a monotonic background. A closer
investigation leads to the conclusion that this
background is very well described by the same
parabolic dependence as the one which we just
discussed for the mesoscopic line~\cite{VVM95} (see
the solid line in Fig.~\ref{MesLine}a). As long as
the width of the strips $w$, forming the loop, is
much smaller than the loop size, the total shift of
$T_c(H)$ can be written as the sum of an oscillatory
part and the monotonic background given by
Eq.~(\ref{TCBLine})~\cite{VVM95,Gro68}:
\begin{equation}
T_c(H)=T_{c0} \left[ 1 - \frac{\pi^2}{3} \left(
\frac{w \: \xi(0) \mu_0 H}{\Phi_0} \right)^2 - \frac{\xi^2(0)}{R^2}
\left( n - \frac{\Phi}{\Phi_0} \right)^2 \right]
\label{TCBRing}
\end{equation}
where $R^2=R_1 \: R_2$ is the product of inner and
outer loop radius, and the magnetic flux threading
the loop $\Phi=\pi R^2 \mu_0 H$. The integer $n$ has
to be chosen so as to maximize $T_c(H)$ or, in other
words, selecting $E_{LLL}(H)$.

The LP~oscillations originate from the fluxoid
quantization requirement, which states that the
complex order parameter $\Psi_s$ should be a
single-valued function when integrating along a
closed contour
\begin{equation}
\oint \vec{\nabla} \varphi \cdot dl = n \: 2 \pi
\; \; \; \; \; \; \; \; \; \; \; \; \; \; \; \;
n=\cdots \: , -2,-1,0,1,2,  \: \cdots
\label{Fluxoid}
\end{equation}
where we have introduced the order parameter $\Psi_s
= |\Psi_s| \exp \: (\imath \varphi)$. Fluxoid
quantization gives rise to a circulating
supercurrent in the loop when $\Phi \neq n
\Phi_0$, which is periodic with the applied flux
$\Phi / \Phi_0$.

Using the sample dimensions and the value for
$\xi(0)$ obtained for the mesoscopic line (with the
same width $w=0.15 \: \mu m$), the $T_c(H)$ for the
loop can be calculated from Eq.~(\ref{TCBRing})
without any free parameter. As shown in
Fig.~\ref{MesLine}b, the agreement with the
experimental data is very good.

Another interesting feature of the mesoscopic loop
or other structures is the unique possibility they
offer for studying nonlocal effects~\cite{StrN96}.
In fact, a single loop can be considered as a 2D
artificial quantum orbit with a {\it fixed radius},
in contrast to Bohr's description of atomic
orbitals. In the latter case the stable radii are
found from the quasiclassical quantization rule,
stating that only an integer number of wavelengths
can be set along the circumference of the allowed
orbits. For a superconducting loop, however,
supercurrents must flow, in order to fulfill the
fluxoid quantization requirement
(Eq.~(\ref{Fluxoid})), thus causing oscillations of
$T_c$ versus $H$.

In order to measure the resistance of a mesoscopic
loop, electrical contacts have, of course, to be
attached to it, and as a consequence the confinement
geometry is changed. This "disturbing" or "invasive"
aspect can now be exploited for the study of
nonlocal effects~\cite{StrN96}. Due to the
divergence of the coherence length $\xi(T)$ at $T =
T_{c0}$ (Eq.~(\ref{XiT})) the coupling of the loop
with the attached leads is expected to be very
strong for $T
\rightarrow T_{c0}$.

\begin{figure}[h]
\centerline{\psfig{figure=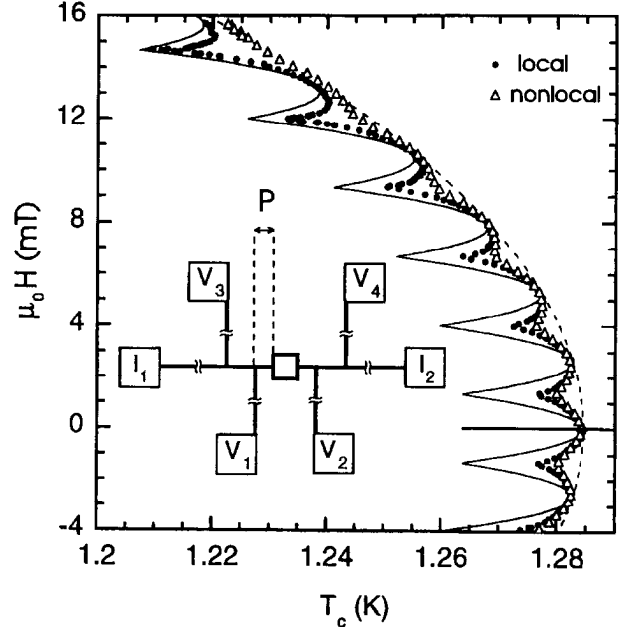}}
% \vspace{8.5cm}
\caption{Local ($V_1/V_2$) and nonlocal phase boundary
($V_1/V_3$ or $V_2/V_4$) measurements. The transport
current flows through $I_1/I_2$. The solid and
dashed lines correspond to the theoretical $T_c(H)$
of an isolated loop and a one-dimensional line,
respectively. The inset shows a schematic of the
mesoscopic loop with various contacts ($P=0.4\:\mu
m$).}
\label{NL1}
\end{figure}

Fig.~\ref{NL1} shows the results of these
measurements. Both "local" (potential probes
$V_1/V_2$ across the loop) and "nonlocal" (potential
probes $V_1/V_3$ or $V_2/V_4$ aside of the loop)
LP~oscillations are clearly observed. For the
"local" probes there is an unexpected and pronounced
increase of the oscillation amplitude with
increasing field, in disagreement with previous
measurements on Al microcylinders~\cite{Gro68}. In
contrast to this, for the "nonlocal" LP~effect the
oscillations rapidly vanish when the magnetic field
is increased.

When increasing the field, the background
suppression of $T_c$ (Eq.~(\ref{TCBLine})) results
in a decrease of $\xi(T)$. Hence, the change of the
oscillation amplitude with $H$ is directly related
to the temperature-dependent coherence length. As
long as the coherence of the superconducting
condensate extends over the nonlocal voltage probes,
the nonlocal LP~oscillations can be observed.

The importance of an "arm" attached to a mesoscopic
loop was already demonstrated theoretically by de
Gennes in 1981~\cite{dGA81}. For a perfect 1D loop
(vanishing width of the strips) adding an "arm" will
result in a decrease of the LP~oscillation
amplitude, what we indeed observed at low magnetic
fields, where $\xi(T)$ is still large. With these
experiments, we have proved that adding probes to a
structure considerably changes both the confinement
topology and the phase boundary $T_c(H)$.

The effect of topology on $T_c(H)$, related to the
presence of the sharp corners in a square loop, has
been considered by Fomin {\it et
al.}~\cite{FominSSC97,FominPRB}. In the vicinity of
the corners the superconducting condensate sustains
a higher applied magnetic field, since at these
locations the superfluid velocity is reduced, in
comparison with the ring. Consequently, in a
field-cooled experiment, superconductivity will
nucleate first around the corners~\cite{FominSSC97}.
Eventually, for a square loop, the introduction of a
{\it local} superconducting transition temperature
seems to be needed. As a result of the presence of
the corner, the $H_{c3}(T)$ of a wedge with an angle
$\theta$~\cite{Fomin98} will be strongly enhanced at
the corner resulting in the ratio $H_{c3}/H_{c2}
\approx 3.79$ for $\theta \approx 0.44 \,
\pi$~\cite{Fomin98}.

\subsection{Dot structure}

\begin{figure}[h]
\centerline{\psfig{figure=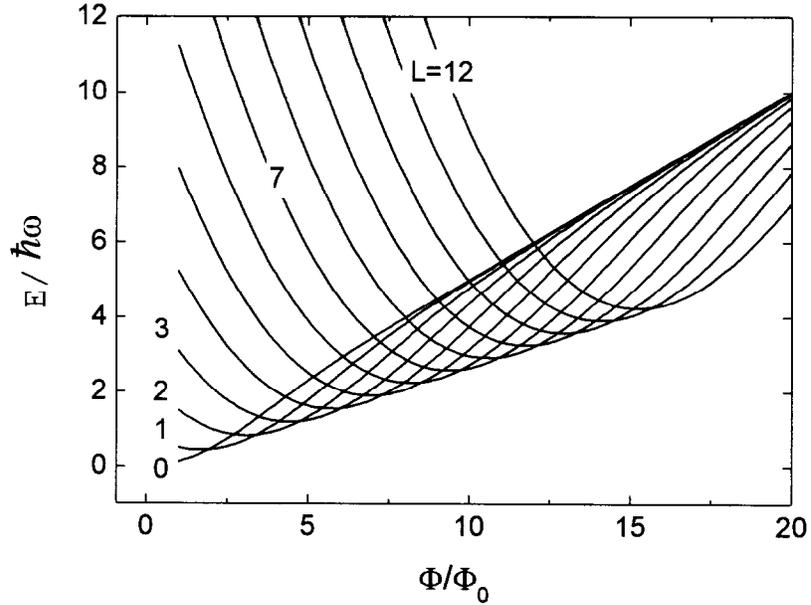}}
% \vspace{8.2cm}
\caption{Energy level scheme versus normalized flux
$\Phi / \Phi_0$ for a superconducting cylinder in a
magnetic field parallel to the axis. The cusp-like
$H_{c3}(T)$ line is formed due to the change of the
orbital quantum number $L$.}
\label{DotCalc}
\end{figure}

The Landau level scheme for a cylindrical dot with
"superconducting" boundary conditions
(Eq.~(\ref{GLBound})) is presented in
Fig.~\ref{DotCalc}. Each level is characterized by a
certain orbital quantum number $L$ where $\Psi_s =
|\Psi_s| \exp \: (\mp \imath L
\varphi)$~\cite{PME96}. The levels, corresponding to
the sign "+" in the argument of the exponent are not
shown since they are situated at energies higher
than the ones with the sign "-". The lowest Landau
level in Fig.~\ref{DotCalc} represents a cusp-like
envelope, switching between different $L$ values
with changing magnetic field. Following our main
guideline that $E_{LLL}(H)$ determines $T_c(H)$, we
expect for the dot the cusp-like superconducting
phase boundary with nearly perfect linear
background. The measured phase boundary $T_c(H)$,
shown in Fig.~\ref{MesLine}b, can be nicely fitted
by the calculated one (Fig.~\ref{DotCalc}), thus
proving that $T_c(H)$ of a superconducting dot
indeed consists of cusps with different
$L$'s~\cite{Bui90}. Each fixed $L$ describes a giant
vortex state which carries $L$ flux quanta $\Phi_0$.
The linear background of the $T_c(H)$ dependence is
very close to the third critical field $H_{c3}(T)
\simeq 1.69
\: H_{c2}(T)$~\cite{Saint-James65}. Contrary to the
loop, where the LP~oscillations are perfectly
periodic, the dot demonstrates a certain
aperiodicity~\cite{VVMScripta}, in very good
agreement with the theoretical
calculations~\cite{Bui90,Benoist}.

The lower critical field of a cylindrical dot
$H_{c1}^{dot}$ corresponds to the change of the
orbital quantum number from $L=0$ to $L=1$, i.e. to
the penetration of the first flux
line~\cite{Benoist}:
\begin{equation}% VERGELIJKING57
\mu_0 H_{c1}^{dot}= 1.924 \, \frac{\Phi_0}{\pi \, R^2} \, .
\label{BeZw}
\end{equation}

For a long mesoscopic cylinder described above,
demagnetization effects can be neglected. On the
contrary, for a thin superconducting disk, these
effects are quite
essential~\cite{Deo,Schweigert,Palacios}. For a
mesoscopic disk, made of a Type-I superconductor,
the phase transition between the superconducting and
the normal state is of the second order if the
expulsion of the magnetic field from the disk can be
neglected, i.e. when the disk thickness is
comparable with $\xi$ and $\lambda$. When the disk
thickness is larger than a certain critical value
first order phase transitions should occur. The
latter has been confirmed in ballistic Hall
magnetometry experiments on individual Al
disks~\cite{GeimNat,GeimAPL,GeimIMEC}. A series of
first order transitions between states with
different orbital quantum numbers $L$ have been seen
in magnetization curves $M(H)$~\cite{GeimNat} in the
field range corresponding to the crossover between
the Meissner and the normal states. Besides the
cusplike $H_{c3}(T)$ line, found earlier in
transport measurements~\cite{VVM95,Bui90},
transitions between the $L=2$ and $L=1$ states have
been observed~\cite{GeimNat} by probing the
superconducting state below the $T_c(H)$ line with
Hall micromagnetometry. Still deeper in the
superconducting area the recovery of the normal
$\Phi_0$-vortices and the decay of the giant vortex
state might be expected~\cite{Palacios}. The former
has been considered in Ref.~\cite{BuzBrison94} in
the London limit, by using the image method.
Magnetization and stable vortex configurations have
been recently analyzed in mesoscopic disks in
Refs.~\cite{Deo,Schweigert,Palacios}.

\section{Conclusions}
We have carried out a systematic study of and
quantization phenomena in submicron structures of
superconductors. The main idea of this study was to
vary the boundary conditions for confining the
superconducting condensate by taking samples of
different topology and, through that, to modify the
lowest Landau level $E_{LLL}(H)$ and therefore the
critical temperature $T_{c}(H)$. Different types of
individual nanostructures were used: line, loop and
dot structures. We have shown that in all these
structures, the phase boundary $T_{c}(H)$ changes
dramatically when the confinement topology for the
superconducting condensate is varied. The induced
$T_{c}(H)$ variation is very well described by the
calculations of $E_{LLL}(H)$ taking into account the
imposed boundary conditions. These results
convincingly demonstrate that the phase boundary in
$T_{c}(H)$ of mesoscopic superconductors differs
drastically from that of corresponding bulk
materials. Moreover, since, for a known geometry
$E_{LLL}(H)$ can be calculated a priori, the
superconducting critical parameters, i.e.
$T_{c}(H)$, can be controlled by designing a proper
confinement geometry. While the optimization of the
superconducting critical parameters has been done
mostly by looking for different materials, we now
have a unique alternative - to improve the
superconducting critical parameters of {\it the same
material} through the optimization of {\it the
confinement topology} for the superconducting
condensate and for the penetrating magnetic flux.
\\

\noindent
{\it Acknowledgements}---The authors would like to
thank V.~Bruyndoncx, E.~Rosseel, L.~Van~Look,
M.~Baert, M.~J.~Van~Bael, T.~Puig, C.~Strunk,
A.~L\'{o}pez, J.~T.~Devreese and V.~Fomin for
fruitful discussions and R.~Jonckheere for the
electron beam lithography. We are grateful to the
Flemish Fund for Scientific Research (FWO), the
Flemish Concerted Action (GOA) and the Belgian
Inter-University Attraction Poles (IUAP) programs
for the financial support.

\label{lastpage}

\end{document}